# Enhanced Stability and Linearly Polarized Emission from CsPbI$_3$ Perovskite Nanoplatelets through A-site Cation Engineering

Woo Hyeon Jeong[1,2], Junzhi Ye[2]*, Jongbeom Kim[3], Rui Xu[4], Xinyu Shen[1,5], Chia-Yu Chang[2], Eilidh L. Quinn[2], Myoung Hoon Song[3], Peter Nellist[6], Henry J. Snaith[5], Yunwei Zhang[4], Bo Ram Lee[1]* and Robert L. Z. Hoye[2]*

[1] School of Advanced Materials Science and Engineering, Sungkyunkwan University, Suwon, 16419, Republic of Korea

[2] Inorganic Chemistry Laboratory, University of Oxford, Oxford, OX1 3QR, United Kingdom

[3] Department of Materials Science and Engineering, Ulsan National Institute of Science and Technology (UNIST), Ulsan, 44919 Republic of Korea

[4] School of Physics, Sun Yat-sen University, Guangzhou, 510275 China

[5] Clarendon Laboratory, Department of Physics, University of Oxford, Oxford, OX1 3PU United Kingdom

[6] Department of Materials, University of Oxford, Oxford OX1 3PH United Kingdom

*Correspondence: junzhi.ye@chem.ox.ac.uk (J.Y.), brlee@skku.edu (B.R.L.), robert.hoye@chem.ox.ac.uk (R.L.Z.H.)

# ABSTRACT

The anisotropy of perovskite nanoplatelets (PeNPLs) opens up many opportunities in optoelectronics, including enabling the emission of linearly polarized light. But the limited stability of PeNPLs is a pressing challenge, especially for red-emitting $CsPbI_3$. Herein, we address this limitation by alloying FA into the perovskite cuboctahedral site. Unlike Cs/FA alloying in bulk thin films or nonconfined nanocubes, FA incorporation in nanoplatelets requires meticulous control over the reaction conditions, given that nanoplatelets are obtained in kinetically-driven growth regimes instead of thermodynamically-driven conditions. Through *in-situ* photoluminescence (PL) measurements, we find that excess FA leads to uncontrolled growth, where phase-impurities and nanoplatelets of multiple thicknesses co-exist. Restricting the FA content to up to 25% Cs substitution enables monodisperse PeNPLs, and increases the PL quantum yield (from 53% to 61%), exciton lifetime (from 18 ns to 27 ns), and stability in ambient air (from ~2 days to >7 days) compared to $CsPbI_3$. This arises due to hydrogen bonding between FA and the oleate and oleylammonium ligands, anchoring them to the surface to improve optoelectronic properties and stability. The reduction in non-radiative recombination, improvement in the nanoplatelet aspect ratio, and higher ligand density lead to FA-containing PeNPLs more effectively forming edge-up superlattices, enhancing the PL degree of linear polarization from 5.1% ($CsPbI_3$) to 9.4% ($Cs_{0.75}FA_{0.25}PbI_3$). These fundamental insights show how the stability limitations of PeNPLs could be addressed, and these materials grown more precisely to improve their performance as polarized light emitters, critical for utilizing them in next-generation display, bioimaging and communications applications.

# INTRODUCTION

Colloidal lead halide perovskite nanocrystals (PNCs) have emerged as one of the most promising materials candidates for display applications, owing to their narrow spectral bandwidth, high defect tolerance, and exceptional luminescence properties, such as near unity photoluminescence quantum yield (PLQY)[1-8]. Moreover, the bandgap of PNCs can easily be tuned by varying their composition and size due to quantum and dielectric confinement. The ability to precisely control the synthesis of these materials enables the shape and size of these nanocrystals to be tailored across different dimensionalities: 0D quantum dots, 1D nanorods, 2D nanoplatelets, and 3D nanocubes[9-11]. Among these, perovskite nanoplatelets (PeNPLs), which are a few unit cells thick and strongly confined only in the out-of-plane direction, exhibit appealing properties, including high exciton binding energy and exciton fine structure splitting[6]. Exciton fine structure splitting enables the production of linearly polarized light, making these PeNPLs excellent candidates for optical applications requiring polarization, for example to enhance the efficiency and contrast from light-emitting diode (LED) displays. Other benefits from linearly polarized light include improving directionality in lasers, improved resolution in bioimaging, enabling optical encoding in anti-counterfeit labels, enhancing selectivity in photodetectors, and providing polarization control in quantum light sources[12-18].

Despite these excellent optoelectronic properties, there remain important challenges with PeNPLs. One of these is maintaining high PeNPL monodispersity in colloidal solution or after depositing to form thin films, which is usually adversely affected by aggregation or surface-reconstruction-induced merging. This undesirable behaviour can broaden the emission profile, or lead to multiple distinct peaks from electronically-isolated PeNPLs with different thicknesess[9,19]. The phase and conformational transitions present in $CsPbI_3$ further limits their stability. The black phases ($α, β$, and $γ$) of $CsPbI_3$, which are intrinsically photoactive, are less stable at room temperature compared to the yellow $δ$-phase[20-24]. The ionic radius of the $Cs^+$ cation is insufficiently large to ideally fit in the cuboctahedral sites in the cubic perovskite structure, such that there is facile phase transition to $δ$-$CsPbI_3$[25,26].

Recently, a variety of strategies have been reported to improve the stability and optical properties of $CsPbI_3$ PeNPLs. Organic ligands, such as ammonium halides and phosphonic acids, have been used to control the synthesis and passivation of PeNPLs, demonstrating the formation of well-aligned $CsPbI_3$ PeNPLs[27,28]. Similarly, divalent metal cation dopants (e.g., $Mn^{2+}$, $Zn^{2+}$) have been incorporated into $CsPbI_3$ PeNPLs to enhance structural stability and achieve well-ordered, low-aspect-ratio PeNPLs[26,29,30]. However, although these approaches have enhanced the colloidal stability of red-emitting PeNPLs, the fundamental issue of achieving phase-stable NCs in thin film form in air remains unresolved, leading to phase transitions occurring in a few days[26,27,30].

On the other hand, in the case of PNCs, several studies have addressed the structural instability of $CsPbI_3$ by incorporating formamidinium ($FA^+$) cations, which have larger ionic radii than $Cs^+$. This A-site cation (Cs and FA) alloying has been shown to stabilize the perovskite structure while suppressing optical losses in PNCs[31-38]. In contrast, for PeNPLs, direct synthesis strategies for A-site cation alloys have not yet been reported, as PeNPL synthesis is more sensitive and demands more meticulous control over reaction conditions than PNCs. Achieving structural stabilization of red-emissive PeNPLs via A-site alloying requires a deeper understanding of the synthesis mechanisms, growth processes, and specific properties to successfully develop PeNPLs as promising emitters.

Herein, we rationalize how FA alloying into the A-site of $CsPbI_3$ influences the growth kinetics, and how this in turn influences the shape, uniformity and self-assembly of the nanoplatelets, as well as how FA incorporation affects the structural and environmental stability of the PeNPLs. We use *in-situ* photoluminescence (PL) measurements to understand how the Cs/FA ratio influences the formation of $Cs_{1-x}FA_xPbI_3$ ($x$ = 0, 0.25, 0.5, 0.75) PeNPLs. Our results reveal that $Cs_{1-x}FA_xPbI_3$ with low values of $x$ exhibit low nucleation and growth rates, leading to the formation of well-ordered PeNPLs. In contrast, FA-rich materials exhibit accelerated nucleation and growth kinetics, which induce overgrown and non-uniform PeNPLs. On the other hand, Cs-only PeNPLs demonstrate rapid degradation under ambient conditions, transitioning into the non-photoactive yellow $\delta$-phase $CsPbI_3$. In comparison, Cs/FA alloyed PeNPLs exhibit improved stability in both colloidal solution (PL $t_{50}$ = 210 min at 80 °C) and thin films (maintaining the α-phase over the 7 days) under ambient conditions. The resulting Cs/FA hybrid PeNPLs exhibit a degree of polarization (DOP) of 9.4%, significantly higher than the DOP of 5.1% observed for Cs-only PeNPLs. Our study on the nucleation and growth mechanisms of Cs/FA-based PeNPLs provides valuable insights into their formation process and identifies a promising synthetic pathway for achieving highly stable and efficient linear polarized red-emissive PeNPLs.

# RESULTS

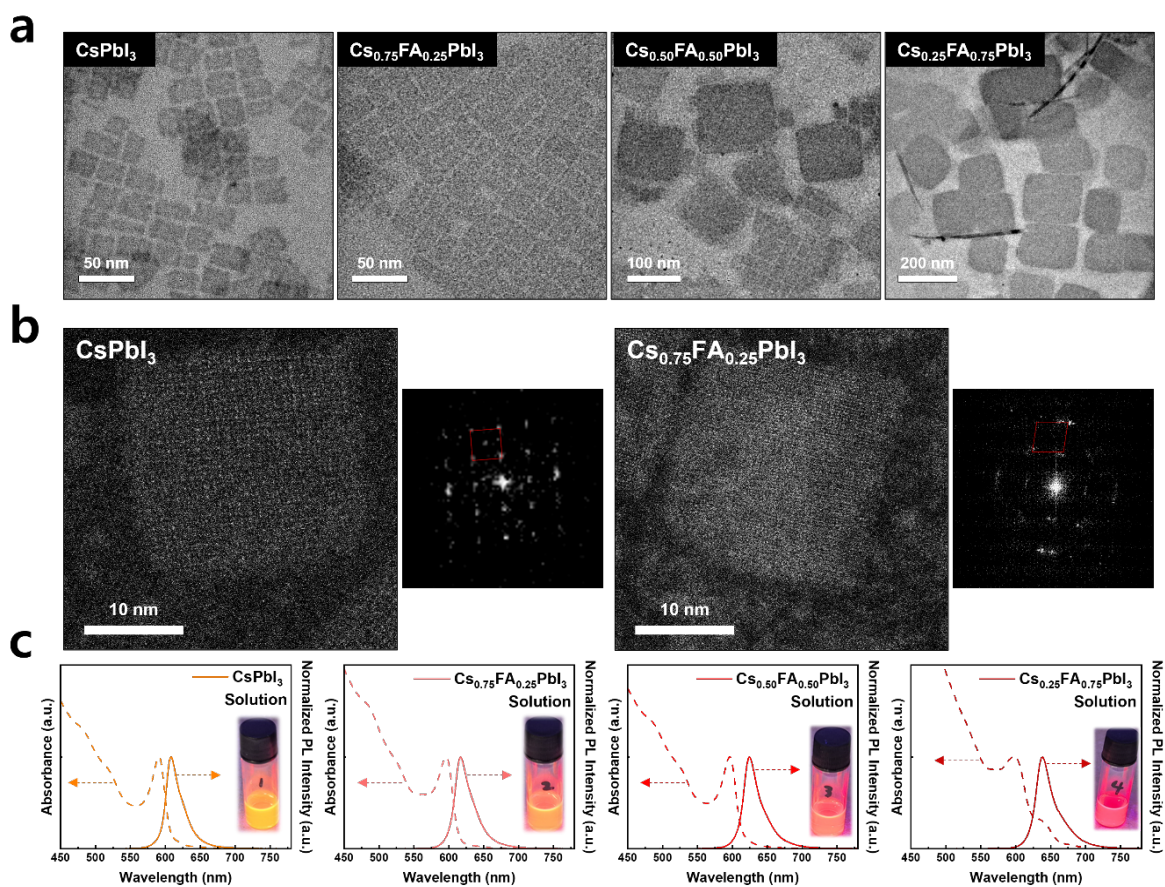

**Fig. 1 | Characterization of Cs/FA alloyed PeNPLs**

**a** High magnification transmission electron microscopy (TEM) images of PeNPLs with different Cs/FA ratios. **b** Dark field scanning TEM images and fast Fourier transform of PeNPLs with 0% and 25% FA incorporation into the A-site. **c** Ultraviolet-visible absorption and photoluminescence spectra of colloidal PeNPL solutions. Inset are photographs of colloidal PeNPL solutions illuminated with a UV lamp (365 nm wavelength).

## Synthesis of (Cs,FA)PbI$_3$ alloy PeNPLs

As a starting point, previous research has demonstrated that the structural stability of Cs$_{1-x}$FA$_x$PbI$_3$ perovskites improves as the FA content increases[33,39,40]. However, these studies focused on bulk perovskites or weakly-confined PNCs, and Cs/FA alloying has not been explored as a strategy to enhance the stability of PeNPLs. Based on these findings, we hypothesized that the stability of well-ordered PeNPLs would improve with increasing FA content. As discussed in the introduction section, PeNPLs require more meticulous synthetic control than isotropic, cube-shaped PNCs because anisotropic nanoplatelets are obtained under kinetically-driven conditions, whereas thermodynamically-governed conditions would favour isotropic nanocubes[9,41]. To investigate the mechanism of nucleation and growth, we synthesized $Cs_{1-x}FA_xPbI_3$ ($x$ = 0, 0.25, 0.50, 0.75) PeNPLs. The $x$ value represents the ratio of Cs-oleate to FA-oleate injected to the $PbI_2$-ligand solution during synthesis (see Methods), and therefore represents the nominal Cs/FA ratio. Transmission electron microscopy (TEM) images of each PeNPL sample are shown in Fig. 1a and Fig. S1 (Supplementary Information). The PeNPLs were dispersed in different solvents and drop-cast onto grids to control their edge-up (hexane) and face-down (octane) orientation. The evaporation rate, which is determined by the vapour pressure of the alkane solvent the PeNPLs are redispersed in after purification, affects the thermodynamic and kinetic processes that control the orientation of the nanoplatelets during film formation[12,16]. TEM analysis indicates that $x$ = 0 and $x$ = 0.25 PeNPLs exhibit a uniform two-dimensional platelet morphology with consistent dimensions, where the variation in the length of the PeNPLs was <13% of the median length. Fig. 1b shows the dark-field scanning transmission electron microscopy (STEM) image and fast Fourier transform (FFT) patterns of the $x$ = 0 and $x$ = 0.25 PeNPLs. According to the FFT pattern, the $x$ = 0 PeNPLs show a clear cubic crystal structure, whereas the $x$ = 0.25 PeNPLs show a slightly tilted octahedral crystal structure due to structural distortion caused by the A-site alloying of Cs and FA cation[42-44]. While the thicknesses of these PeNPLs were the same (2.6 ± 0.4 nm), there was an improvement in the regularity of the shape of the PeNPLs with FA alloying. The median lengths of the PeNPLs were the same (26 ± 2 nm) for both compositions, while the widths increased from 21 ± 3 nm for $x$ = 0 to 22 ± 2 nm for $x$ = 0.25, leading to the aspect ratio (AR) reducing from 1.26 ± 0.19 to 1.09 ± 0.06 (Fig. S3, Supplementary Information). Previous studies have reported that PeNPLs with AR values closer to unity have improved stability[26]. We therefore expect that adding a small amount of FA is beneficial (see later for a detailed discussion). However, we found that further increasing the FA content led to overgrowth of the PeNPLs to >100 nm in size. For $x$ = 0.50 PeNPLs, a broad size distribution was observed (Fig. S2, Supplementary Information), ranging from 20 nm to 160 nm, indicating the coexistence of small PeNPLs and overgrown PeNPLs. In contrast, $x$ = 0.75 NPLs were predominantly composed of overgrown PeNPLs exceeding 100 nm.

A strong excitonic peak was observed in the ultraviolet-visible (UV-Vis) absorption spectra of the colloidal PeNPLs (Fig. 1c). The exciton binding energies determined through Elliott model fitting were 212 meV ($x$ = 0), 213 meV ($x$ = 0.25), 165 meV ($x$ = 0.50), and 153 meV ($x$ = 0.75), respectively. For $x$ = 0 and $x$ = 0.25, the exciton binding energies are consistent with previous reports of $CsPbI_3$ PeNPLs with 2.6 nm thickness[16], which falls well below the exciton Bohr diameter (~12 nm)[45]. In the case of $x$ = 0.75 PeNPLs, a relatively weak excitonic peak was observed, which can be attributed to nonuniform growth, with a heterogeneous size distribution caused by the overgrowth of PeNPLs. The PL emission peaks of each colloidal PeNPL solution were observed to be centred at 608, 616, 624, and 638 nm. These are all consistent with 3-monolayer thick PeNPLs, albeit with a slight red-shift in PL with increasing FA content in the A-site (Fig. 1c), consistent with an overall reduction in the degree of confinement.

Subsequently, X-ray diffraction (XRD) measurements were conducted for each PeNPL composition to investigate structural changes resulting from changing the FA content in the A-site of the perovskites. According to the XRD patterns (Fig. S5, Supplementary Information), increasing the FA cation content induces larger lattice parameters, which shifts the diffraction peaks to smaller angles[33,35,46]. Furthermore, the XRD patterns of the $x = 0$ and $x = 0.25$ PeNPLs exhibit superlattice peaks, arising from diffraction from stacked PeNPLs arranged face down. Notably, the overall XRD peak intensities gradually decrease with increasing FA content, which can be attributed to the overgrowth of PeNPLs, leading to larger lateral dimensions and a heterogeneous size distribution, both of which contribute to more random stacking orientations.

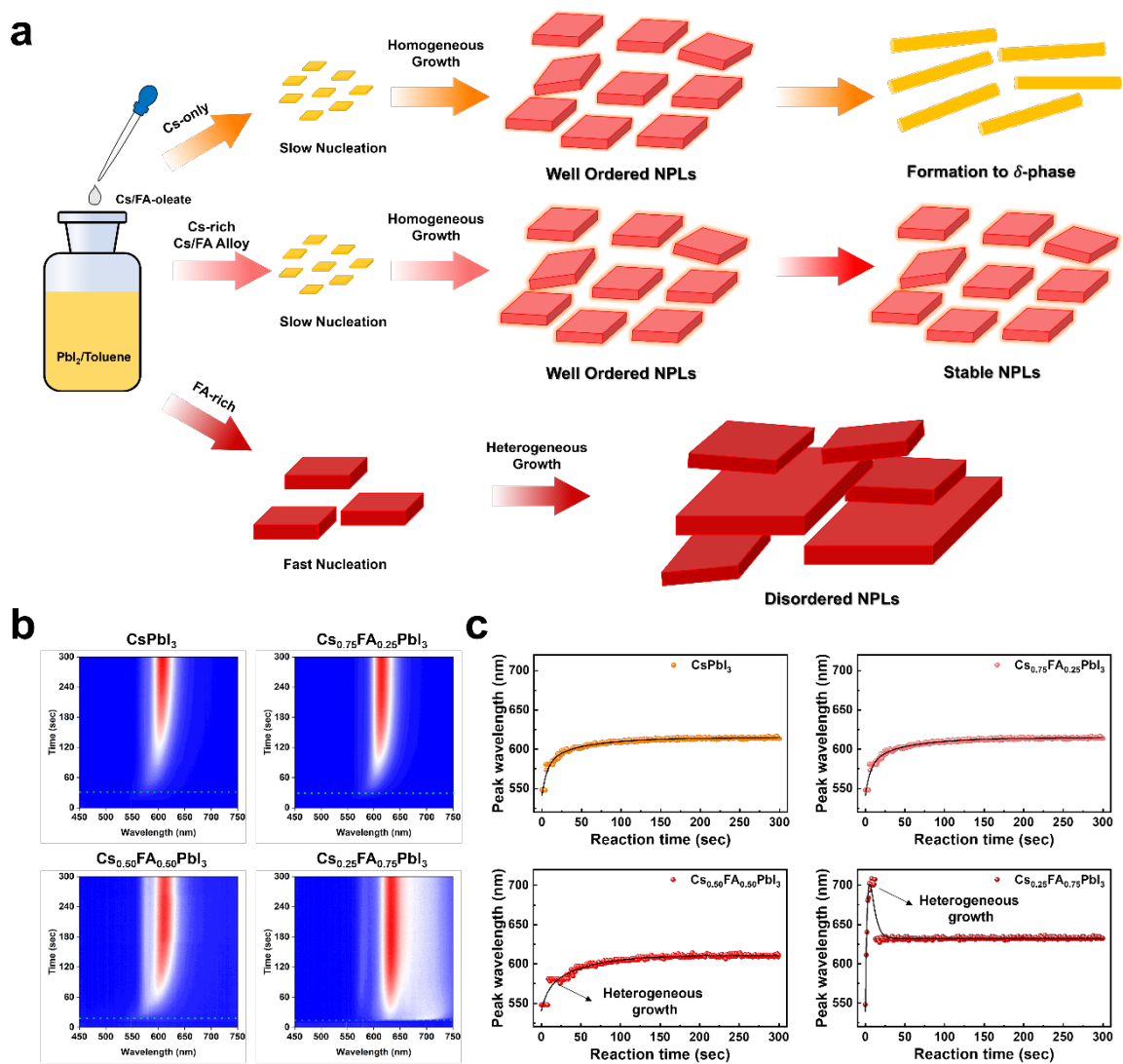

**Fig. 2 | Nucleation and growth kinetics of PeNPLs**

**a** Schematic illustration of the synthesis of PeNPLs with tunable Cs/FA ratios. **b** *In-situ* PL spectra during the formation of each PeNPL in colloidal solution. **c** Evolution of the PL peak wavelength over time after injecting the Cs/FA oleate solution into the reaction mixture. Data points were collected at 1 s time intervals.

## Growth kinetics of Cs/FAPbI$_3$ alloy PeNPLs

Forming a well-ordered PeNPL superlattice requires uniform PeNPLs. Our results in Fig. S5 (Supplementary Information) show that this is accomplished with low FA contents, but not when the fraction of FA in the A-site exceeds 0.5. This is due to increased disorder, and less uniform sizes and shapes of the PeNPLs for high FA content. To explain this, we propose a mechanism that is illustrated in Fig. 2a. Specifically, we propose that $x = 0$ and $x = 0.25$ PeNPLs undergo slow nucleation with little structural distortion, leading to the formation of well-ordered PeNPLs through homogeneous growth. Conversely, we propose that faster nucleation takes place with increasing FA content, and that structural distortions are induced by FA cations, which have a different ionic radius than Cs$^+$, disrupting the uniform growth of PeNPLs. This results in the formation of disordered PeNPLs with different sizes, which we describe as heterogeneous growth.

To experimentally test these hypotheses, we monitored the change in the PL spectra of the colloidal solution, illuminated with a 405 nm wavelength continuous wave (cw) laser, over time after injecting the Cs-oleate/FA-oleate precursor into the PbI$_2$-ligand solution. We could assign the thickness of the PeNPLs obtained based on the PL peak wavelengths[16,18]. According to the *in-situ* PL spectra shown in Fig. 2b and Fig. S6 (Supplementary Information), we observed significant differences in the rates of nucleation and growth for each composition. For $x = 0$ PeNPLs, nucleation occurs approximately 30 s after the injection of Cs-oleate and starts with $n = 2$ nanoplatelets (~550 nm, depending on the Cs/FA composition), followed by growth to the $n = 3$ nanoplatelets (~610 nm) within 2 min (Fig. 2c). In the case of $x = 0.25$ PeNPLs, the growth kinetics are similar to those of $x = 0$ PeNPLs, with slightly faster nucleation and growth but still exhibiting overall homogeneous growth, in which PeNPLs with different thicknesses or perovskite nanocubes are not formed. In contrast, for compositions with $x \geq 0.5$, different growth behavior was observed. For $x = 0.50$ PeNPLs, the PL peak positions over time reveal rapid formation of $n = 2$ (~550 nm) within the initial reaction period (~30 s), followed by slow growth to $n = 3$ nanoplatelets (~600 nm) (Fig. 2c). For FA-rich compositions, an intriguing phenomenon was observed during the growth of $x = 0.75$ PeNPLs. PL peaks centered at ~700 nm (likely corresponding to FAPbI$_3$ PNCs) were detected within the first 10 s. Following this, the PL peak shifted to 631 nm within 8 s (Fig. 2c), which we attribute to FAPbI$_3$ nanocrystals changing to (Cs,FA)PbI$_3$ PeNPLs. However, the final

PL of these PeNPLs is the most red-shifted out of all samples compared. For $x = 0.50$ and 0.75 PeNPLs, the PL spectra also had additional peaks alongside the main peak corresponding to the $n = 3$ PeNPLs (Fig. 2b and Fig. S6, Supplementary Information), indicating the co-existence of both low-$n$ PeNPLs and perovskite nanocubes. These non-uniform species could hinder the formation of well-ordered PeNPL superlattices.

To further substantiate our understanding of the effects of FA on nanoplatelet growth, we investigated the nucleation and growth kinetics of FAPbI$_3$ PeNPLs under equivalent synthesis conditions (Fig. S7, Supplementary Information). *In-situ* PL for these colloidal nanoplatelet solutions demonstrate nearly instantaneous nucleation upon injection, with almost all PeNPLs becoming $n = 3$ within the first 10 s (Fig. S7c, Supplementary Information). This behaviour contrasts sharply with the slower nucleation and growth kinetics observed for CsPbI$_3$ PeNPLs, providing compelling evidence to support the hypothesis that FA addition to CsPbI$_3$ led to faster nucleation and growth. Among the compositions studied, CsPbI$_3$ and Cs$_{0.75}$FA$_{0.25}$PbI$_3$ PeNPLs emerged as the most promising candidates, providing optimal structural and optical properties, such as well-ordered PeNPLs with a narrow size distribution and the highest exciton binding energies which will be the focus of further investigations.

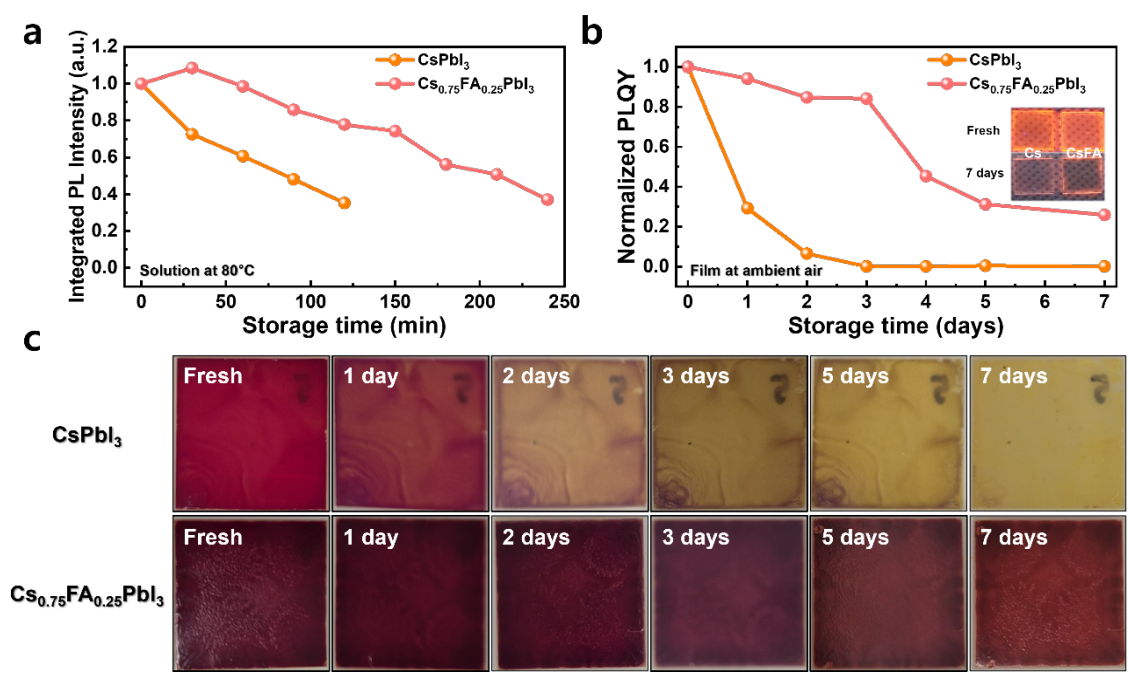

**Fig. 3 | Optical and phase stability of PeNPLs**

**a** Integrated PL intensity of colloidal PeNPL solutions as a function of time stored at 80 °C in ambient air. **b** Change in the normalized PLQY of PeNPL thin films. **c** Photographs of thick PeNPL films as function of time stored in ambient air.

## Enhanced phase and optical stability of PeNPLs

Based on the understanding of the synthesis and growth kinetics of PeNPLs, we investigate the effect of FA incorporation on the stability of PeNPLs. As a starting point, we heated a colloidal solution of $CsPbI_3$ and $Cs_{0.75}FA_{0.25}PbI_3$ to an elevated temperature (80 °C) and monitored the PL intensity over time (Fig. 3a and Fig. S8, Supplementary Information). As shown in Fig. 3a, the PL intensity of the $x = 0$ PeNPLs rapidly decreases, reaching 50% of the original PL intensity ($t_{50}$) after 90 min. For the $x = 0.25$ PeNPLs, the $t_{50}$ increased to 210 min. The decrease in PL intensity over time with heating is due to the detachment of the labile organic ligands from the surface of the PeNPLs, and the degradation of the perovskite to the yellow $\delta$-phase. These results therefore suggest that FA addition reduces ligand detachment.

To measure the stability of the PeNPLs when assembled together to form films, we prepared thin films by spin coating onto glass substrates, as well thick films by drop casting the colloidal solution onto glass substrates. We monitored the stability of these films in air under ambient conditions (40% relative humidity, 20 °C, storage in the dark). For the thin films, we monitored the PLQY over time, and found that the $x = 0$ PeNPLs lost 90% of its initial PLQY within 2 days, however, the $x = 0.25$ PeNPL film maintained 84% of the original PLQY over the course of 3 days, and finally lost 70% of its initial PLQY after 7 days (Fig. 3b). To monitor the phase stability, we measured changes in the diffraction pattern of the thick films stored in air under the same conditions (Fig. 3c and Fig. S9, Supplementary Information). For the $x = 0$ PeNPLs, the yellow $\delta$-phase starts to appear after just 1 day, after which the film rapidly becomes the $\delta$-phase after 3 days. The $x = 0.25$ PeNPL films were more stable, mostly maintaining the α-phase over the 7-day testing period, with only traces of $\delta$-phase appearing from day 3 onwards.

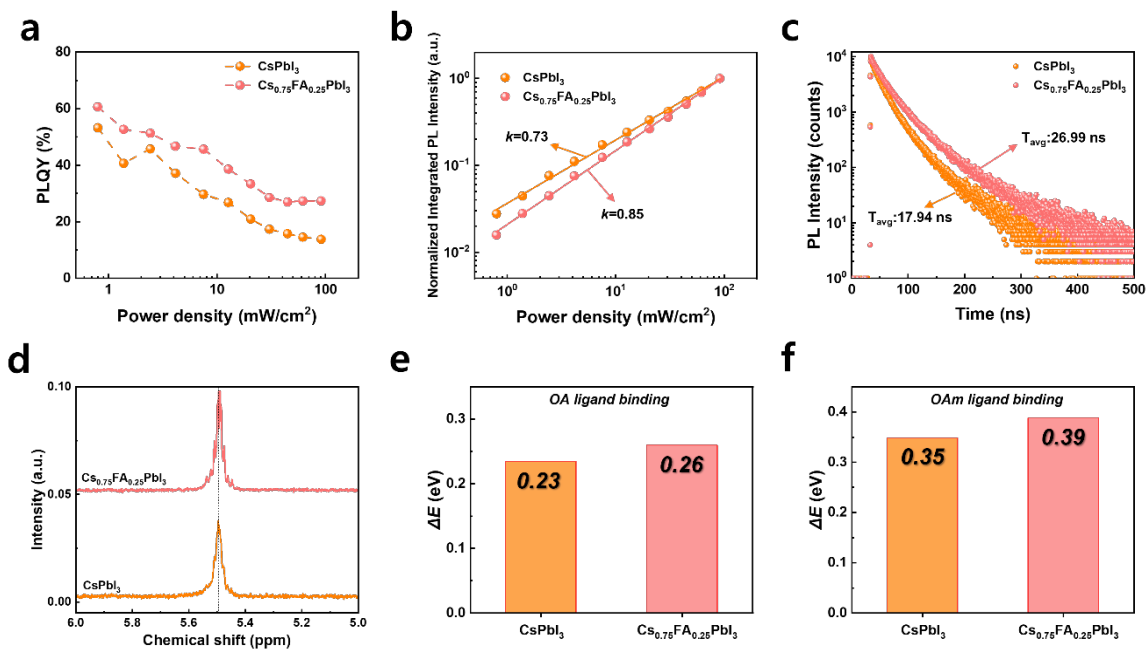

**Fig. 4 | Optical properties and surface chemistry of PeNPLs**

Power density-dependent **a** PLQY, **b** integrated PL peak intensity and **c** time-resolved PL decay curve of colloidal PeNPLs. **d** $^1$H-NMR spectra of colloidal PeNPLs dispersed in deuterated toluene. DFT-calculated binding energy of **e** oleic acid and **f** oleylamine on the surface of each PeNPLs surface.

**Effect of FA on ligand binding strength to perovskite surface**

To understand the influence of FA alloying on the optoelectronic properties of PeNPLs, we conducted excitation power dependent PLQY and time-resolved photoluminescence (TRPL) measurements (Fig. 4a-c). As shown in Fig. 4a, at all excitation power densities, the PLQY values of $x = 0.25$ PeNPLs were higher than those of $x = 0$ PeNPLs. The PLQY decreased with increasing excitation power density in both cases (refer to Fig. S10, Supplementary Information for PL spectra). Given the high exciton binding energies (refer to earlier section on Elliott model fitting), this decrease in PLQY with increasing excitation power density is due to bimolecular exciton-exciton annihilation. Fitting a power law function to the PL intensity as a function of excitation power density yields exponentials close to unity (Fig. 4b). This is consistent with excitonic behavior, whereas a sublinear dependence would typically indicate the involvement of non-radiative pathways such as trap-assisted recombination or exciton-exciton annihilation[16]. We fit the PL decay curves with a phenomenological multi-exponential decay function (Fig. 4c). The weighted average of the time constants fit to quantitatively describe the PL decays ($\tau_{avg}$) increased from 17.94 ns ($x = 0$) to 26.99 ns ($x = 0.25$). There was therefore a consistent increase in PLQY and TRPL $\tau_{avg}$ from $x = 0$ to $x = 0.25$, suggesting that the enhancement in $\tau_{avg}$ is due to fewer non-radiative recombination processes with a small amount of FA incorporation to the PeNPLs.

The surface ligands (here oleic acid OA, and oleylamine OAm) play a key role in enhancing the optoelectronic properties and stability of PeNPLs through the suppression of surface defects. These ligands also play a critical role on long-term stability by reducing the exposure of the surface of the perovskite to moisture and oxygen, which can cause PeNPL degradation[48-50]. We analyzed the ligand density on the PeNPLs through solution phase $^1$H-NMR spectroscopy (Fig. 4d; full spectra in Fig. S11, Supplementary Information). We obtained the NMR spectrum from 3 mg of PeNPLs and 0.5 mg of ferrocene dispersed in 1 mL of deuterated toluene (d-toluene) solution. As shown in Fig. 4d, the integration of the NMR peaks was performed with respect to the ligand peak at a chemical shift of around 5.5 ppm, which corresponds to the vinyl peaks (C=C) from both the oleic acid and oleylamine ligands[52]. From this, we found that the $x = 0.25$ PeNPLs had about 1.2 times higher ligand density than the $x = 0$ PeNPLs. In addition, a slight shift of the ligand peak from 5.49 ppm to 5.48 ppm was observed in the $x = 0.25$ PeNPLs compared to $x = 0$ PeNPLs. This is consistent with the formation of hydrogen bonding between the FA cation and the ligands, leading to the increase in ligand density observed for these PeNPLs, and is consistent with observations previously reported for perovskite NCs with FA cation[33,51].

To gain a deeper understanding of the binding properties of each ligand (OA, OAm) on the PeNPL surface, we performed density functional theory (DFT) calculations to determine the binding energies of these ligands with $x = 0$ and $x = 0.25$ perovskites. The DFT-calculated binding energies for each perovskite surface and ligand are presented in Fig. 4e-f, and the slab model of the perovskite structure is shown in Fig. S12, Supplementary Information. The results indicate that FA cations on the PeNPL surface are involved in hydrogen bonding with the functional groups of the ligands, as evidenced by the increased binding energies of OA$^-$ and OAm$^+$ ligands on the $x = 0.25$ PeNPLs compared to the $x = 0$ PeNPLs (Fig. 4e-f). The enhanced ligand density and stronger surface binding energy between the ligands and PeNPLs account for the reduced non-radiative recombination observed in Cs/FA alloyed PeNPLs. These findings support the hypothesis that $x = 0.25$ PeNPLs enable stronger ligand passivation on the PeNPLs surface as well as structural thermodynamic stability, and suggest a pathway to achieve highly emissive, long-term stable PeNPLs.

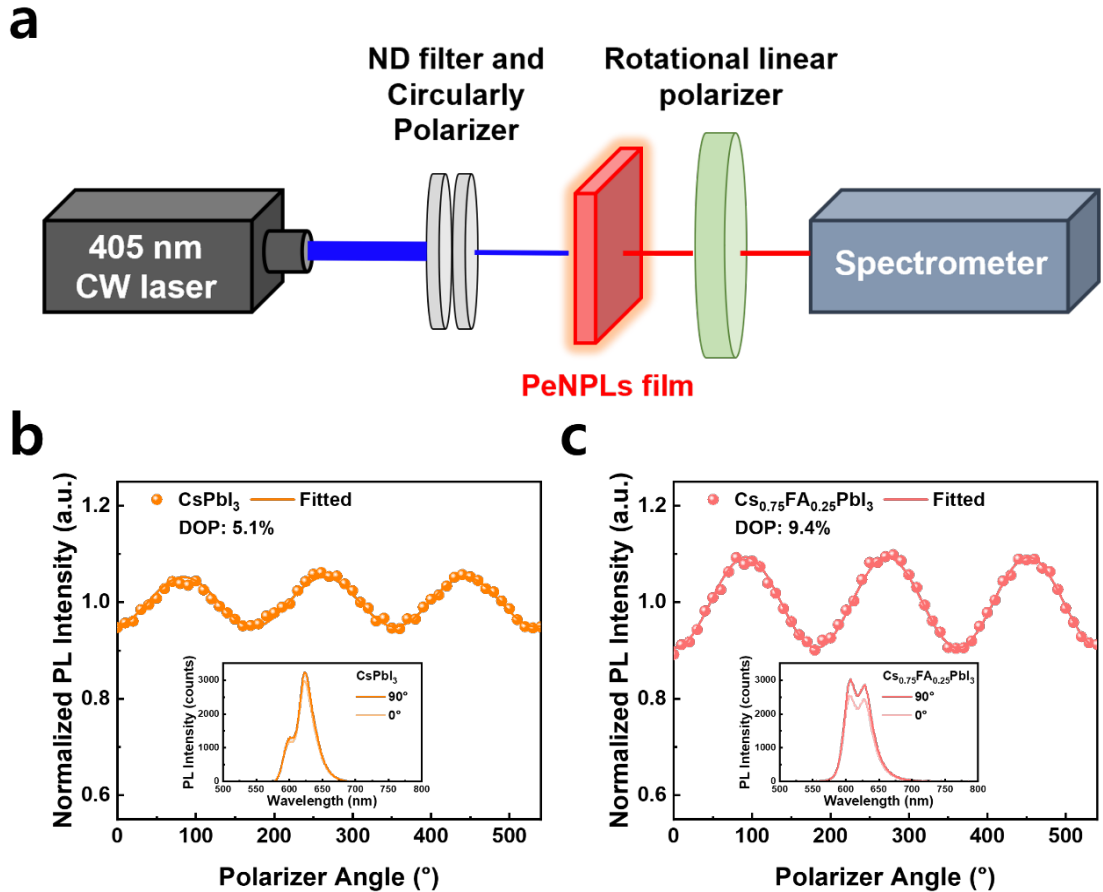

**Fig. 5 | Linearly polarized emission of PeNPL films**
**a** Experimental setup for measuring the degree of linear polarization (DOP) in PL. **b** Polarization dependence of the normalized PL intensity of $CsPbI_3$ PeNPL film. **c** Polarization dependence of the normalized PL intensity of $Cs_{0.75}FA_{0.25}PbI_3$ PeNPLs film. Inset are the PL spectra as a function of angle.

**Linearly-polarized photoluminescence from PeNPLs films**

As outlined in the introduction, light emitters with strongly confined and anisotropic structures, particularly two-dimensional PeNPLs, can exhibit linearly polarized emission, making them highly promising for numerous optical applications. Based on the successful synthesis and control over the uniformity, stability and optoelectronic properties of the PeNPLs obtained in this work, we measured the degree of polarization (DOP) from ensembles of these PeNPLs in films made by drop-casting. The experimental setup for DOP measurements is shown in Fig. 5a. The PeNPL film was fabricated on a glass substrate by drop-casting a colloidal PeNPL solution that had hexane as the solvent. The rapid evaporation enabled PeNPLs to be kinetically trapped in the edge-up orientation. As shown in Fig. 5b and 5c, the PeNPLs exhibited a maximum DOP of 5.1%, with a median value of 4.3% across 6 samples for $x$ = 0 PeNPLs. This increased to 9.4% (median value: 9.0% across 6 samples) for $x$ = 0.25 PeNPLs. Histograms showing the distribution in DOP values measured are shown in Fig. S13 and S14, Supplementary Information.

The enhanced DOP for $x$ = 0.25 can be attributed to three factors. Firstly, the improvement in surface ligand density, as discussed earlier, likely results in denser PeNPL arrangement during superlattice formation, which enhances linear polarization properties. Secondly, the reduction in non-radiative recombination losses, enabled by the effective defect passivation through Cs/FA alloying, contributes to a higher fraction of radiative exciton recombination. Since exciton recombination is inherently influenced by the anisotropic crystal structure of PeNPLs, reducing trap-assisted recombination enables the intrinsic polarization of excitons to emerge more prominently, leading to improved linear polarization properties. Thirdly, as shown inset in Fig. 5b and c, the $x$ = 0.25 PeNPLs experienced reduced agglomeration than the $CsPbI_3$ PeNPLs when assembled to form thin films, such that there was a larger contribution in their emission from the more strongly-confined $n$ = 3 PeNPLs. This would lead to enhanced exciton fine structure splitting, which is needed to achieve a higher DOP. Overall, these enhanced linear polarization properties of PeNPLs highlight their potential as efficient and stable red emitters for next-generation display technologies.

## DISCUSSION

In this work, we demonstrated that A-site hybridization through Cs/FA alloying significantly enhances the phase stability and optical properties of red-emitting I-based PeNPLs. By investigating the nucleation and growth kinetics, we show that low FA alloying results in slower nucleation and homogeneous growth, forming well-ordered structures, while FA-rich PeNPLs exhibit rapid nucleation and heterogeneous growth, leading to structural disorder. The optimized Cs/FA PeNPLs showed superior phase stability in thin films under ambient conditions, maintaining their photoactive phase for over seven days, compared to the rapid degradation observed in Cs-only PeNPLs. Additionally, Cs/FA alloying enhanced the interaction between surface ligands and the PeNPLs surface through hydrogen bonding, resulting in increased ligand density and improved radiative recombination. As a result, the Cs/FA alloy PeNPLs exhibited a 1.8-fold improvement in their emission of linearly polarized light, with a degree of polarization of 9.4%, compared to 5.1% for Cs-only PeNPLs. This enhancement suggests that Cs/FA alloy PeNPLs could serve as promising materials for applications requiring polarized light emission, such as optical communication, quantum emitters and optoelectronic devices.

## DATA AVAILABILITY

Raw data for the main text and supplementary information available from the Oxford Research Archive repository [DOI to be made available before publication]

## CONFLICT OF INTERSET

The authors declare no competing interests.

## ACKNOWLEDGEMENTS

We thank to Dr. Mohsen Danaie and the Diamond Light Source for access and support in the use of the electron Physical Science Imaging Centre (Instrument E02, proposal no. MG40059-2) that contributed to the results presented here. J. Y. and R. L. Z. H. thank the UK Research and Innovation (UKRI) for funding through a Frontier Grant (no. EP/X022900/1), awarded via the 2021 ERC Starting Grant scheme. R. L. Z. H. is funded through a Science & Technology Facilities Council / Royal Academy of Engineering Senior Research Fellowship (no. RCSRF2324-18-68). C.-Y. C. thanks the Oxford-Taiwan Graduate Scholarship and Clarendon Fund Scholarship. E. Q. acknowledges funding from the EPSRC Centre for Doctoral Training in Inorganic Chemistry for Future Manufacturing (OxICFM; no. EP/S023828/1). Y. W. Z. acknowledges funding from the National Key R&D Program of China No. 2023YFA1610000, National Natural Science Foundation of China under Grant No.12304036, the Guangdong Basic and Applied Basic Research Foundation (2023A1515010071), and the Fundamental Research Funds for the Central Universities, Sun Yat-sen University (23xkjc016). R. X. acknowledges funding from Fundamental Research Funds for the Central Universities, Sun Yat-sen University Grant Code 74130-31610059. This work was supported by the National Research Foundation of Korea (NRF) grant funded by the Korean government (MSIT) (RS-2025-00523067, NRF-2022R1A2C4002248, 2021M3H4A1A02049006 and RS-2025-00516815). This work was partly supported by Korea Institute for Advancement of Technology (KIAT) grant funded by the Korean Government (MOTIE) (RS-2024-00418086, HRD Program for Industrial Innovation).

## AUTHOR CONTRIBUTION

W.H.J., J.Y., B.R.L. and R.L.Z.H. conceptualized this work. W.H.J and J.K. performed formal analysis under supervision of M.H.S., B.R.L. and R.L.Z.H. R.X. and Y.Z. performed DFT calculation. X.S., H.J.S. performed optical analysis. C.-Y.C. and P.N. performed TEM measurement. E.Q. performed NMR measurement. W.H.J. drafted the first version of the manuscript. J.Y., B.R.L., R.L.Z.H. reviewed and revised the manuscript. This work was carried out under supervision of B.R.L. and R.L.Z.H. All the authors contributed to the discussion of the manuscript.

# SUPPLEMENTARY INFORMATION

Supplementary information accompanies the manuscript on the Light: Science & Applications website (http://www.nature.com/lsa)